\documentclass[]{raa}
\usepackage{graphicx,times}
\usepackage{amsmath}
\usepackage{natbib}
\usepackage{amssymb,amsmath}
\bibpunct{(}{)}{;}{a}{}{,}

\usepackage[a4paper=true,pagebackref=true]{hyperref}

\hypersetup{pdftitle = The title of my PDF, pdfauthor = My name, pdfsubject= The subject, pdfkeywords = keyword1 keyword2 keyword3}
\hypersetup{colorlinks = true, linkcolor = green, anchorcolor = red, citecolor = blue, filecolor = red, pagecolor = red, urlcolor = red}

\begin{document}
 \title{How common can the origin of clusters Cha\,I, Cha\,II, $\epsilon$~Cha and $\eta$~Cha be?}
 \volnopage{ {\bf 20XX} Vol.\ {\bf X} No. {\bf XX}, 000--000}
   \setcounter{page}{1}
   \author{V. V. Bobylev, A. T. Bajkova}
   \institute{Pulkovo Astronomical Observatory, St.-Petersburg 196140, Russia;
   {\it bob-v-vzz@rambler.ru}\\G
   \vs \no
   {\small Received 20XX Month Day; accepted 20XX Month Day}
 }

\abstract{
The trajectories of the clusters $\eta$~Cha, $\epsilon$~Cha and Cha\,I, Cha\,II, constructed backward in time, have been studied. We concluded that the hypothesis about the joint formation of all four of these clusters from one molecular cloud cannot be completely excluded. However, 10--15 Myr ago, all these four clusters were located at approximately the same height above the plane of the Galaxy. Thus, the gas-dust clouds from which all these four clusters were formed were located on one broad front. It is possible that the appearance of the Cha\,I, Cha\,II, $\epsilon$~Cha and $\eta$~Cha clusters may be associated with the impact on such a front of shock waves formed after supernova explosions in the Scorpius-Centauri association. New estimates of the kinematic ages of the clusters Cha~I and Cha\,II are obtained as $0.12\pm0.19$ and $0.05\pm0.15$, respectively. It is shown that the minimum size of the Cha\,I-north and Cha\,I-south clusters corresponded to the time of $0.55\pm0.24$~Myr and $0.04\pm0.18$~Myr ago, respectively, and approximately 1.5~Myr ago the distance between the trajectories of these two groupings was minimal.
}


 \authorrunning{V. V. Bobylev, A. T. Bajkova}
 \titlerunning{How common can the origin of clusters Cha\,I, Cha\,II, $\epsilon$~Cha and $\eta$~Cha be?}
 \maketitle

 \section{Introduction}
Close to the Sun (at a distance of about 100--250 pc), the fourth galactic quadrant is occupied by the Scorpius-Centauri OB association \citep{Blaauw1946,Blaauw1964}. It has three main structural components LCC: (Lower Centaurus Crux), UCL (Upper Centaurus Lupus) and US (Upper Scorpius). A large number of publications have been devoted to the study of the stellar composition and kinematics of the Sco--Cen OB association \citep{Geus1989,Geus1992,Bruijne1999,Zeeuw1999,Sartori2003,Fernandez2008,Poppel2010,Rizzuto2011,Pecaut2012,Pecaut2016,BobBajk2020,BobBajk2023}. The association has a complex morphology and evolutionary history~--- some of its components (LCC, UCL, etc.) have an age of 10--25 Myr, in which the star formation process has already completed. There are also structures (for example, US with an age of $\sim$5~Myr), where star formation is still occurring.

The star formation process in the Sco--Cen OB association generally fits well with the sequential star formation model \citep{Preibish1999}. This model is based on the assumption that shock waves from explosions of massive stars in one part of the association lead to compression of hydrogen clouds in another part, thus triggering the process of star formation.

The velocity field in the Sco--Cen association is very heterogeneous. The study of spatial and kinematic features showed that a whole complex of very young moving stellar groups, small associations and open star clusters is associated with the Sco--Cen OB association, for example, $\rho$~Oph/L1688, $\nu$~Sco, B59, $\eta$~Cha or $\epsilon$~Cha. A detailed summary of more than 30 groupings of various ages, indicating the relationship with the main structural components (LCC, UCL and US) can be found in \cite{Ratzenbock2023}.

It should be noted that the use of data from the Hipparcos project \citep{HIP1997} made it possible to study in detail the features of the Sco--Cen OB association based on bright stars \citep{Zeeuw1999,Sartori2003,Fuchs2006,BobBajk2007}. The implementation of the Gaia project \citep{Gaia Collab2016} made it possible to analyze en masse the faint stars of the Sco--Cen association, which are mostly T~Taurus type stars.

In the work of \cite{Galli2021}, the spatial velocities of clusters $\eta$~Cha, $\epsilon$~Cha and Cha\,I, Cha\,II were compared using data from the Gaia\,DR2 catalog \citep{Gaia Collab2018}. These authors concluded that the spatial motion of $\eta$~Cha, $\epsilon$~Cha is consistent with the spatial motion of stars in the molecular clouds Cha\,I and Cha\,II within 3$\sigma$ of the stated uncertainties in their velocities. Therefore, these clusters may have been formed from the same parent cloud, but further research is needed to confirm this hypothesis. This conclusion of \cite{Galli2021} served as the main impetus for carrying out this study.

The purpose of this work is to analyze the trajectories constructed backward in time of the stars of the clusters $\eta$~Cha, $\epsilon$~Cha and Cha\,I, Cha\,II in order to find out which of them could have formed from a common molecular cloud. Moreover, the analysis of the trajectories of clusters $\eta$~Cha and $\epsilon$~Cha has already been performed by us earlier in the work \cite{BobBajk2024} using data from the Gaia\,DR3 catalog \citep{Gaia Collab2022}.

In this work, the main attention is paid to the clusters Cha\,I and Cha\,II. Candidate stars from the list of \cite{Galli2021} of these clusters we also provided the values of parallaxes and proper motions from the Gaia\,DR3 catalogue. Of particular interest is also the study of the evolutionary relationship between the Cha\,I and Cha\,II clusters.

 \section{Method}
We use a rectangular coordinate system centered on the Sun, where the $x$ axis is directed towards the galactic center, the $y$ axis~--- towards galactic rotation and the $z$ axis~--- to the north pole of the Galaxy. Then $x=r\cos l\cos b,$ $y=r\sin l\cos b$ and $z=r\sin b,$ where $r=1/\pi$ is the heliocentric distance of the star in the kpc, which we calculate through the parallax of the star $\pi$ in mas (milliseconds of arc).

From observations, the radial velocity $V_r$ and two tangential velocity projections $V_l=4.74r\mu_l\cos b$ and $V_b=4.74r\mu_b$ are known, directed along the galactic longitude $l$ and latitude $b$, respectively, expressed in km s$^{-1}$. Here, the coefficient 4.74 is the ratio of the number of kilometers in an astronomical unit to the number of seconds in a tropical year. The components of proper motion $\mu_l\cos b$ and $\mu_b$ are expressed in mas yr$^{-1}$.

Through the components $V_r, V_l, V_b,$ the velocities $U,V,W,$ are calculated, where the velocity $U$ is directed from the Sun to the center of the Galaxy, $V$ in the direction of rotation of the Galaxy and $W$ to the north galactic pole:
 \begin{equation}
 \begin{array}{lll}
 U=V_r\cos l\cos b-V_l\sin l-V_b\cos l\sin b,\\
 V=V_r\sin l\cos b+V_l\cos l-V_b\sin l\sin b,\\
 W=V_r\sin b                +V_b\cos b.
 \label{UVW}
 \end{array}
 \end{equation}
Thus, the velocities $U,V,W$ are directed along the corresponding coordinate axes $x,y,z$.

To plot the orbits of stars in the coordinate system rotating around the center of the Galaxy, we use the epicyclic approximation~\citep{Lindblad1927}:
 \begin{equation}
 \renewcommand{\arraystretch}{1.8}
 \begin{array}{lll}\displaystyle
 x(t)= x_0+{U_0\over \displaystyle \kappa}\sin(\kappa t)+     \\
      +{\displaystyle V_0\over \displaystyle 2B}(1-\cos(\kappa t)),  \\
 y(t)= y_0+2A \biggl(x_0+{\displaystyle V_0\over\displaystyle 2B}\biggr) t-\\
       -{\displaystyle \Omega_0\over \displaystyle B\kappa} V_0\sin(\kappa t)
       +{\displaystyle 2\Omega_0\over \displaystyle \kappa^2} U_0(1-\cos(\kappa t)),\\
 z(t)= {\displaystyle W_0\over \displaystyle \nu} \sin(\nu t)+z_0\cos(\nu t),
 \label{EQ-Epiciclic}
 \end{array}
 \end{equation}
where $t$ is time in million years (based on the ratio pc/Myr=0.978 km s$^{-1}$),
$A$ and $B$ are Oort constants; $\kappa=\sqrt{-4\Omega_0 B}$ is epicyclic frequency; $\Omega_0$ is angular velocity of galactic rotation of the local standard of rest, $\Omega_0=A-B$; $\nu=\sqrt{4\pi G\rho_0}$ is frequency of vertical oscillations, where $G$ is gravitational constant, and $\rho_0$ is stellar density in the near-solar neighborhood.

The parameters $x_0,y_0,z_0$ and $U_0,V_0,W_0$ in the system of equations~(\ref{EQ-Epiciclic}) denote the current positions and velocities of the stars, respectively. The elevation of the Sun above the galactic plane $h_\odot$ is assumed to be $16$~pc according to the work~\cite{BobBajk2016}. The velocities $U,V,W$ are calculated relative to the local standard of rest using the values $(U_\odot,V_\odot,W_\odot)=(11.1,12.2,7.3)$~km s$^{-1}$ obtained by \cite{Schonrich2010}. We took $\rho_0=0.1~M_\odot/$pc$^3$~\citep{Holmberg2004}, which gives $\nu=74$~km s$^{-1}$ kpc$^{-1}$. We use the following values of the Oort constants $A=16.9$~km s$^{-1}$ kpc$^{-1}$ and $B=-13.5$~km s$^{-1}$ kpc$^{-1}$, close to modern estimates. An overview of such estimates can be found, for example, in the works of \cite{Krisanova2020} or \cite{Donlon2024}.

\begin{figure}[t]
{ \begin{center}
  \includegraphics[width=0.8\textwidth]{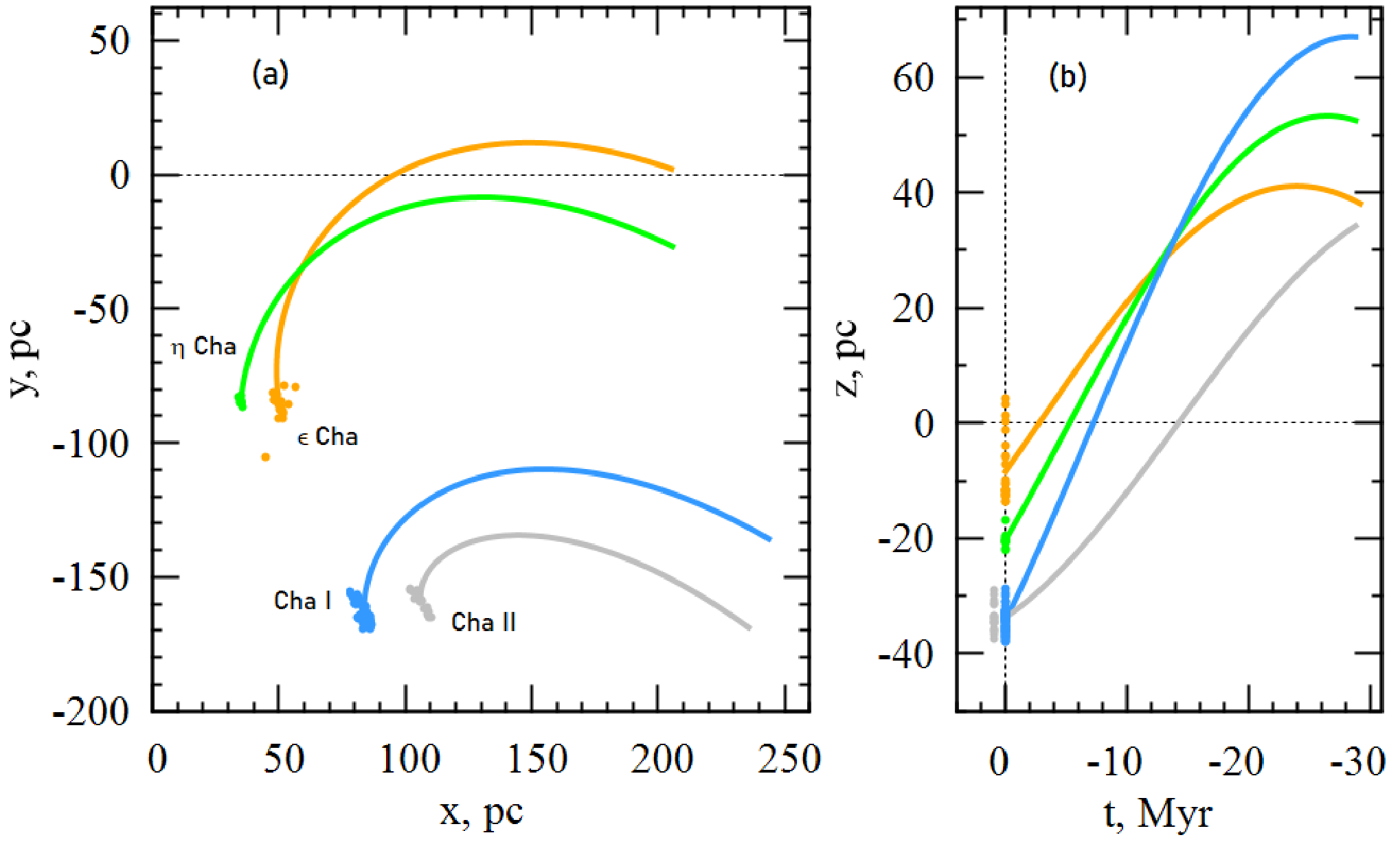}
  \caption{
Distribution of four clusters in projection onto the $xy$ plane and their trajectories integrated backward in time over an interval of 30~Myr~(a), their distribution and trajectories in the vertical direction~(b).}
 \label{f1-not Gal Rot}
\end{center}}
\end{figure}
\begin{figure}[t]
{ \begin{center}
   \includegraphics[width=0.99\textwidth]{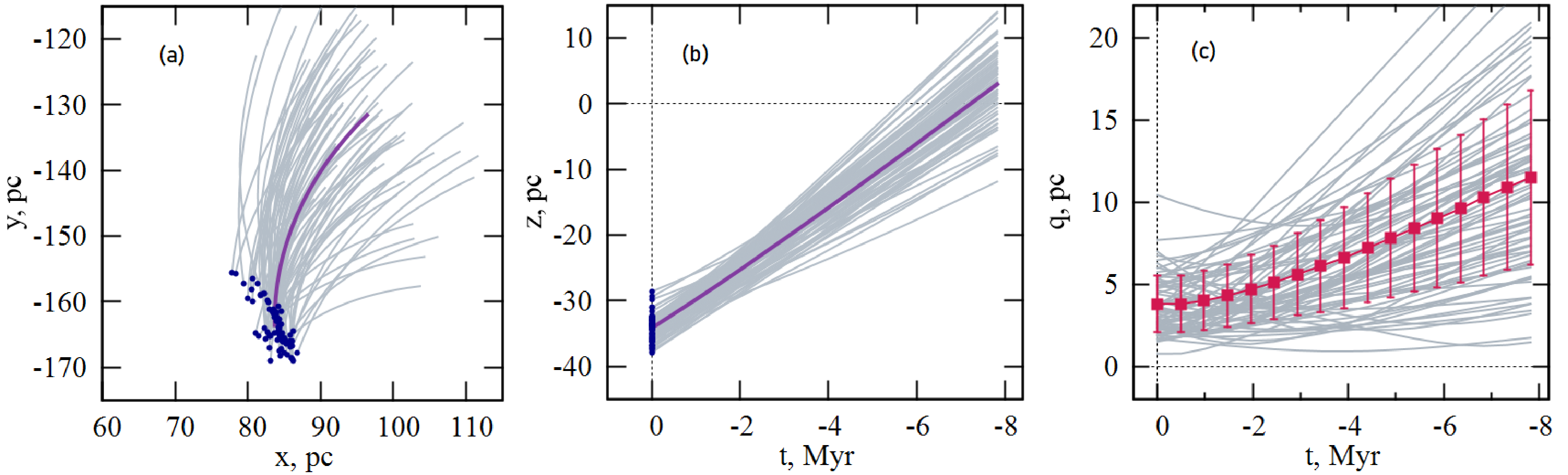}
  \caption{
Distribution of 76 Cha~I stars in projection onto the $xy$ plane and their trajectories integrated backward in time over an interval of 8~Myr~(a), their distribution and trajectories in the vertical direction~(b), dependence of the parameter $q$ ( deviation from the trajectory of the kinematic center) for each star versus time, the averaged interval values with the corresponding dispersions are shown in red ~(c), the dark thick line in panels (a) and (b) shows the trajectory of the kinematic center.}
 \label{f2-Cha-I-all}
\end{center}}
\end{figure}
\begin{figure}[t]
{ \begin{center}
   \includegraphics[width=0.99\textwidth]{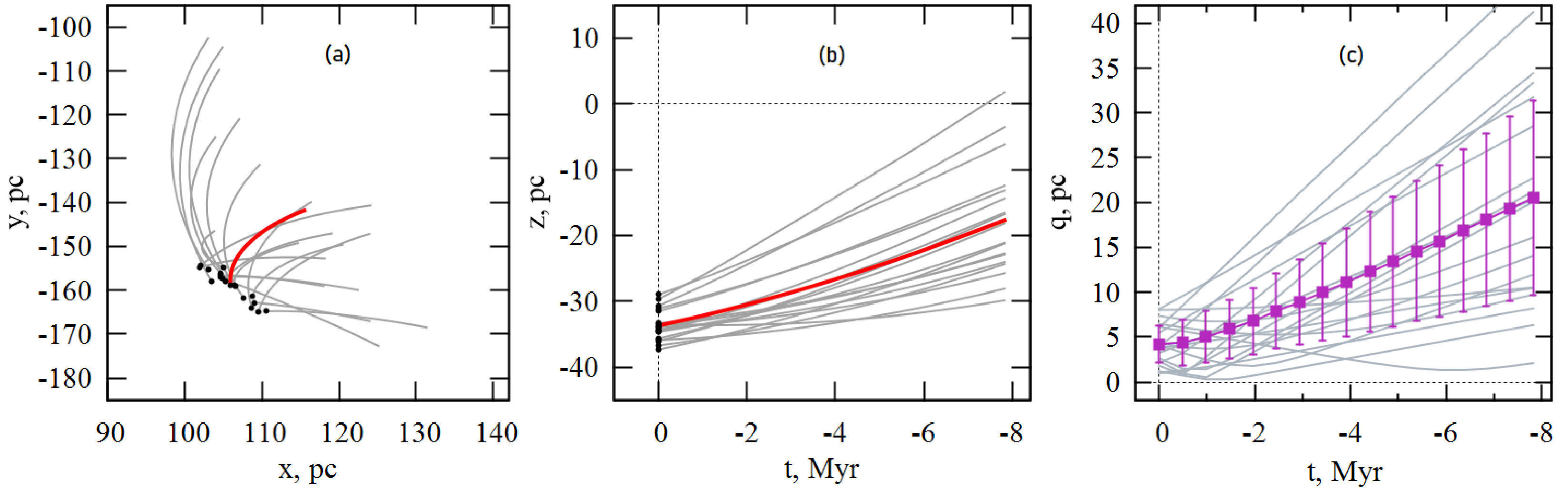}
  \caption{
Distribution of 19 Cha~II stars in projection onto the $xy$ plane and their trajectories integrated backward in time over an interval of 8~Myr~(a), their distribution and trajectories in the vertical direction~(b), dependence of the parameter $q$ ( deviation from the trojectory of the kinematic center) for each star versus time, the averaged interval values with the corresponding dispersions are shown in purple ~(c), the dark thick line in panels (a) and (b) shows the trajectory of the kinematic center.}
 \label{f3-Cha-II-all}
\end{center}}
\end{figure}

\section{Data}
In this work, the analysis uses the list of probable members of the Cha\,I and Cha\,II clusters from the work of \cite{Galli2021}. Moreover, only stars with radial velocity measurements were taken. The selected stars were provided with parallax and proper motion values from the Gaia\,DR3 catalogue.

The radial velocity values from the Gaia\,DR3 catalog, available for a number of stars, were not used in this work. The fact is that the ground-based radial velocities of the stars of both clusters Cha\,I and Cha\,II, collected in the work of \cite{Galli2021}, have values close to 15~km s$^{-1}$ with low dispersion. But the radial velocities of the stars of these clusters in the Gaia\,DR3 catalog, if measured, in many cases have negative values and large measurement errors.

\section{Results and discussion}
\subsection{$\eta$~Cha, $\epsilon$~Cha and Cha\,I, Ch\,II}
Fig.~\ref{f1-not Gal Rot} shows the distribution of stars of four clusters $\epsilon$~Cha, $\eta$~Cha, Cha\,I and Cha\,II in projection onto the $xy$ plane and their trajectories integrated backward in time over an interval of 30~Myr, their distribution and trajectories in the vertical direction are also given.

To set the trajectory of the kinematic center of the Cha\,I cluster, the average positions and velocities were calculated for 76~stars:
${\overline x}_0=83.7\pm0.7$~pc,
${\overline y}_0=-163.6\pm0.6$~pc,
${\overline z}_0=-50.2\pm1.2$~pc,
${\overline U}_0=-11.0\pm0.6$~km s$^{-1}$,
${\overline V}_0=-19.3\pm1.8$~km s$^{-1}$ and
${\overline W}_0=-11.5\pm0.5$~km s$^{-1}$. The average distance to the cluster was obtained as ${\overline r}=190.5\pm0.4$~pc.

Using the found values of the average coordinates and velocities, the trajectory of the kinematic center was constructed. Based on the coordinate differences between the star and the kinematic center $\Delta x,\Delta y,\Delta z$ at each moment of integration for each star, we calculate the value of the prarameter $q$ of the following form:
\begin{equation}
 q=\sqrt{\Delta x^2+\Delta^2+\Delta z^2},
 \label{qq}
 \end{equation}
which characterizes the deviation of the star from the trajectory of the kinematic center. Note that the trajectories of the stars are calculated taking into account the elevation of the Sun above the galactic plane. Thus, in all our figures, the coordinate $z$ reflects the position of the stars relative to the plane of the Galaxy.

In the work of \cite{Galli2021} for the Cha\,I cluster using 78~stars with Gaia\,DR2 parallaxes, the following average velocities were calculated
${\overline U}_0=-11.0\pm0.5$~km s$^{-1}$,
${\overline V}_0=-19.5\pm0.8$~km s$^{-1}$,
${\overline W}_0=-11.2\pm0.3$~km s$^{-1}$. The value of the average distance to the cluster Cha\,I, ${\overline r}=189.4^{+0.8}_{-0.7}$~pc, was obtained by these authors from 160 stars (Table ~3 of their work).

For the Cha\,II cluster of 19~stars, we found the following values:
${\overline x}_0=105.9\pm1.0$~pc,
${\overline y}_0=-158.8\pm4.1$~pc,
${\overline z}_0=-49.7\pm0.6$~pc
${\overline U}_0=-10.9\pm1.0$~km s$^{-1}$,
${\overline V}_0=-18.1\pm1.5$~km s$^{-1}$ and
${\overline W}_0=-8.6\pm0.8$~km s$^{-1}$, and ${\overline r}=197.3\pm1.1$~pc as well.

In \cite{Galli2021}, the following average velocities were calculated for the Cha\,II cluster of 19~stars using Gaia\,DR2 parallaxes:
${\overline U}_0=-11.0\pm2.9$~km s$^{-1}$,
${\overline V}_0=-18.1\pm4.2$~km s$^{-1}$,
${\overline W}_0=-8.5\pm1.4$~km s$^{-1}$ and average distance
${\overline r}=197.5^{+1.0}_{-0.9}$~pc calculated from 31 stars (Table.~3 of their works). Thus, we have a good agreement between us and \cite{Galli2021} in estimating the average distances and velocities of clusters Cha\,I and Cha\,II.

Let's note some interesting points. Firstly, in Fig.~\ref{f1-not Gal Rot}(a) the elongation along the line of sight of the stars in the Cha\,I and Cha\,II clusters is clearly visible. The effect is clearly related to the variance of errors in estimates of distances to stars. For example, the size of Cha\,I in the transverse direction is about 5~pc, and along the line of sight its size is about 15~pc. Assuming that all these stars belong to one spherically symmetric cluster, we obtain an independent estimate of the error in the distance to the cluster $\sim5\%.$

Secondly, as can be seen from Fig.~\ref{f1-not Gal Rot}(a), in the $xy$ plane in the past, the trajectories of only two clusters ~--- $\epsilon$~Cha and $\eta$~Cha intersect about 8~Myr ago. In this plane, the trajectories Cha\,I and Cha\,II run parallel to each other. There is no convergence between the trajectories of the groupings Cha\,I, Cha\,II and $\epsilon$~Cha, $\eta$~Cha. Thus, for the joint formation of clusters Cha\,I, Cha\,II, $\epsilon$~Cha and $\eta$~Cha from one parent gas-dust cloud, it is required that it be a GMC (Giant Molecular Cloud) larger than 100~pc. Such a scenario cannot be completely excluded. For example, according to the work of \cite{Meingast2021}, many young open clusters in the solar vicinity have extensive coronae (with a radius of up to 100 pc) formed together with the cluster core. This suggests that the star formation region may be quite extensive.

An overview of the problems of the origin and evolution of molecular clouds, GMCs, in particular, can be found in the work of \cite{Williams2000}. An analysis of the sizes of young stellar associations and associated molecular clouds in the solar neighborhood with a radius of less than 3.5~kpc is given in \cite{Zhou2024}, from which it is clear that the diameter value of 200~pc for such structures lies at the very edge of the distribution, and the maximum (typical size) is less than 50~pc.

Thirdly, as can be seen from Fig.~\ref{f1-not Gal Rot}(b), 10--15~million years ago Cha\,I, Cha\,II, $\epsilon$~Cha and $\ eta$~Cha were located at approximately the same height above the Galactic plane. Moreover, three of them were in a very narrow zone (their trajectories intersect in the figure). Thus, the individual gas-dust clouds from which all these four clusters were formed, were located on one broad front. This gives reason to assume that the emergence of the Cha\,I, Cha\,II, $\epsilon$~Cha and $\eta$~Cha clusters may be associated with the impact of shock waves on the front of the clouds formed after supernova explosions in the Sco--Cen association.

As already it is noted in the introduction, the Sco--Cen association identifies three main structures LCC, UCL and US of slightly different ages. The youngest of the three, US ($\sim$5~Myr old), is not suitable for the scenario we propose. But in the associations LCC ($\sim$15~Myr) and UCL ($\sim$10~Myr) explosions of the most massive stars of spectral class O as supernovae should have occurred just about 10~Myr ago. Indeed, at the moment there are no stars of spectral class O in the LCC and UCL associations.

\subsection{Cha\,I and Cha\,II}
Figure~\ref{f2-Cha-I-all} shows the distribution of 76 Cha\,I stars projected onto the $xy$ plane and their trajectories integrated backward in time over an interval of 8~Myr; in the middle panel of the figure is given distribution of Cha~I stars and their trajectories in the vertical direction. Figure~\ref{f2-Cha-I-all}(c) shows the dependence of the parameter $q$ (\ref{qq}) (deviation from the trajectory of the kinematic center) for each star on time. Note that here there are two identical values of the mean and variance for the moments $t=0$ and $t=-0.5$~Myr. This suggests that less than 0.5 million years ago the size of the Cha~I cluster reached its minimum value. That is, the kinematic age of Cha~I is less than 0.5~Myr.

Figure~\ref{f2-Cha-I-all}(c) shows discrete averages. Based on more detailed analysis, we estimated the point in time at which the size of the star system in the past was minimal:
\begin{equation}
t=0.12\pm0.19~\hbox{Myr.}
\label{t-Cha I}
 \end{equation}

Fig.~\ref{f3-Cha-II-all} shows the distribution of 19 Cha\,II stars projected onto the $xy$ plane and their trajectories integrated backward in time over an interval of 8~Myr; the distribution of Cha\,II stars and their trajectories in the vertical direction are given, as well as the dependence of the parameter $q$ for each star on time. In Fig.~\ref{f3-Cha-II-all}(c) the minimum value of the parameter $q$ and the variance is present only at the moment $t=0$~Myr. Based on more detailed calculations for Cha\,II stars, an estimate was obtained for the moment in time at which the size of the stellar system in the past was minimal:
\begin{equation}
t=0.05\pm0.15~\hbox {Myr.}
\label{t-Cha II}
 \end{equation}

The error of the moment $t$ in the results (\ref{t-Cha I}) and (\ref{t-Cha II}) was found as a result of statistical Monte Carlo simulations. It was assumed that the orbits of the stars were constructed with errors of 10\%, distributed according to the normal law.

Thus, we can conclude that there is good agreement between our estimates of the kinematic age of the Cha~I and Cha\,II clusters with isochronous estimates of their age, 1--2~Myr, obtained in the work of \cite{Galli2021}. Moreover, the Cha~I cluster is slightly older than Cha\,II.

 \begin{table}[t]                               
 \caption[]{\small\baselineskip=1.0ex\protect
The average values of coordinates and velocities of Cha\,I-north (top line)
and Cha\,I-south (bottom line).
}
\begin{center}
 \begin{tabular}{|r|r|r|r|r|r|r|}\hline
 \label{t-1}\small
 $n_\star$ & ${\overline x}_0$ & ${\overline y}_0$ & ${\overline z}_0$ &
       ${\overline U}_0$ & ${\overline V}_0$ & ${\overline W}_0$\\
           & pc & pc & pc & km s$^{-1}$ & km s$^{-1}$ & km s$^{-1}$ \\\hline
40 &$84.2\pm0.2$&$-165.2\pm1.0$&$-49.1\pm0.1$&$-10.6\pm0.2$&$-19.3\pm1.6$&$-11.7\pm0.7$ \\
36
&$83.1\pm2.6$&$-161.8\pm3.2$&$-51.4\pm1.6$&$-11.5\pm1.0$&$-19.5\pm2.0$&$-11.3\pm0.6$ \\
  \hline \end{tabular} \end{center}
 \end{table}
\begin{figure}[t]
{ \begin{center}
  \includegraphics[width=0.8\textwidth]{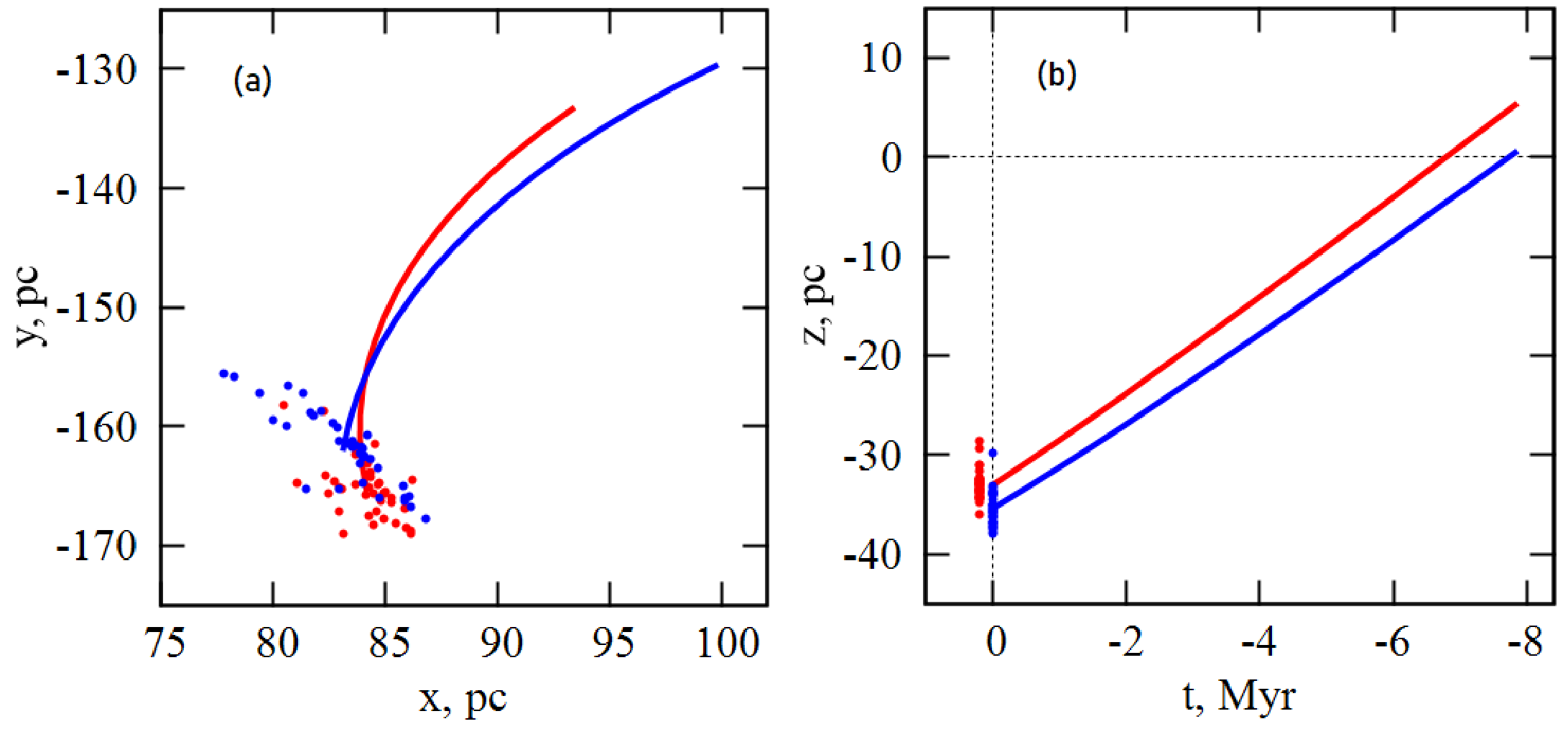}
  \caption{
The distribution of stars of the Cha\,I-north (red dots and lines) and Cha\,I-south (blue dots and lines) clusters in projection onto the $xy$ plane and trajectories of their kinematic centers, integrated backward in time over an interval of 8~million years~(a), the distribution of stars and trajectories of centers in the vertical direction~(b).}
 \label{f4-XY-NS}
\end{center}}
\end{figure}
\begin{figure}[t]
{ \begin{center}
   \includegraphics[width=0.8\textwidth]{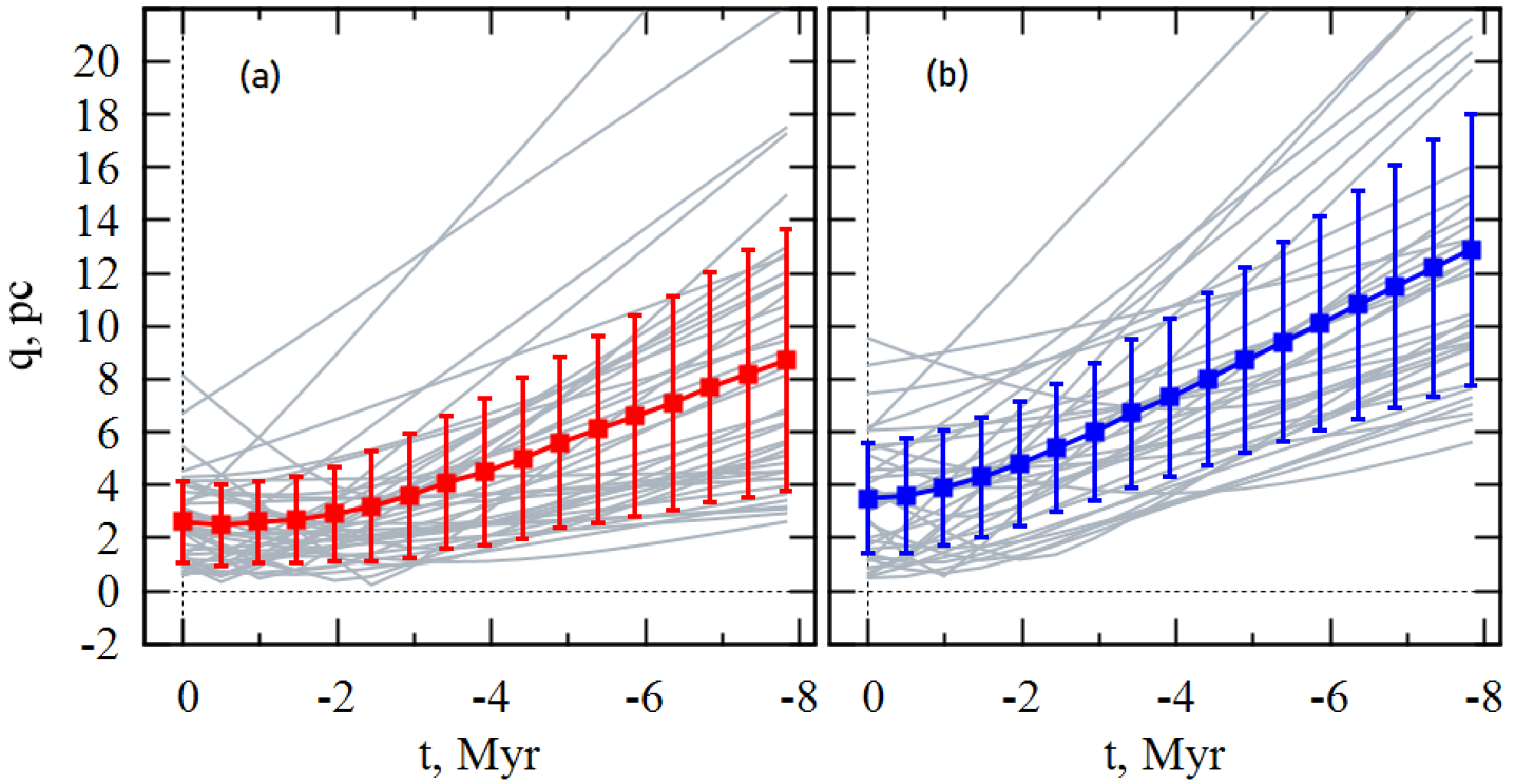}
  \caption{
The dependence of the parameter $q$ on time for stars of the Cha\,I-north cluster with interval averages (red squares and lines) (a) and the Cha\,I-south cluster with interval averages (blue squares and lines) (b).}
\label{f5-NS-q}
\end{center}}
\end{figure}

\subsection{Cha\,I-north and Cha\,I-south}
In \cite{Galli2021}, it is indicated that there may be differences between two structures in the Cha\,I cluster, namely, between Cha\,I-north and Cha\,I-south. Therefore, we repeat our analysis for these two structures.

The average values of the positions and velocities of Cha\,I-north (40 stars each) and Cha\,I-south (36 stars each) are given in the table.~\ref{t-1}. In addition, we found that the average distance to the grouping Cha\,I-north is ${\overline r}=191.8\pm0.4$~pc, and to Cha\,I-north~--- ${\overline r}=189.1\pm0.7$~pc.

The distribution of stars in the Cha\,I-north and Cha\,I-south clusters in projection onto the $xy$ plane and in the vertical direction is given in Fig.~\ref{f4-XY-NS}. The trajectories of the kinematic center of each cluster were calculated using data from the table~\ref{t-1}. As can be seen in panel (a), approximately 1.5~Myr ago the trajectories of the centers of these groups intersected, and in the vertical direction the trajectories ran almost parallel. Thus, in three-dimensional space about 1.5~Myr ago, the distance between the trajectories of these clusters was minimal.

In Fig.~\ref{f5-NS-q} the dependence of the parameter $q$ on time is given for the stars of the clusters Cha\,I-north and Cha\,I-south. The average interval values are also given. Based on a more detailed analysis, a time point was found at which the size of the Cha\,I-north star system was minimal in the past

Figure~\ref{f5-NS-q} shows the dependence of the parameter $q$ on time for stars of the Cha\,I-north and Cha\,I-south clusters. The average interval values are also given.
Based on a more detailed analysis, a time point was found at which the size of the star system Cha\,I-north in the past was minimal
 \begin{equation}
t=0.55\pm0.24~\hbox {Myr,}
 \label{t-ChaI-north}
 \end{equation}
and for Cha\,I-south
 \begin{equation}
t=0.04\pm0.18~\hbox {Myr.}
\label{t-ChaI-south}
 \end{equation}

The $q$ parameter can also be considered as an estimate of the cluster radius. As can be seen from Fig.~\ref{f5-NS-q}(a), the Cha\,I-north cluster is very compact, about 0.5~Myr ago its radius was approximately 5~pc. This is the typical size of a gravitationally bound open star cluster \citep{Piskunov2008,Kharchenko2009}. The radius of the cluster Cha\,I-south is currently about 8~pc. However, it seems that the compression of this complex is still ongoing.

\section*{Conclusion}
A study of the trajectories of the $\eta$~Cha, $\epsilon$~Cha and Cha\,I, Cha\,II clusters, constructed backward in time over a time interval of 30 million years, allows us to conclude the following.

1). In order to form  all these four clusters from a common molecular cloud (the hypothesis of \cite{Galli2021}), the size of such a cloud must be about 100--120 pc. According to current data, such a scenario cannot be completely excluded.

2). However, 10--15~Myr ago, all these four clusters were located approximately at the same height above the plane of the Galaxy. Thus, the gas--dust clouds from which all these four clusters were formed were on the same broad front. It is possible that the occurrence of clusters Cha\,I, Cha\,II, $\epsilon$~Cha and $\eta$~Cha may be associated with the impact of shock waves caused by supernova explosions in the Sco--Cen association on such a front from smaller parent clouds.

3). The kinematics of clusters Cha\,I and Cha\,II have been studied. For this purpose, stars with trigonometric parallaxes and proper motions from the Gaia\,DR3 catalog and radial velocities from the work of \cite{Galli2021} were used, where they were taken from the literature. Moreover, the radial velocities of these stars were measured as a result of ground-based observations. New estimates of the average distance ${\overline r}=190.5\pm0.4$~pc and ${\overline r}=197.3\pm1.1$~pc to clusters Cha\,I and Cha\,II, respectively, are obtained. It is shown that there is good agreement between our estimates of the kinematic age of $t$ obtained for clusters Cha\,I and Cha\,II $0.12\pm0.19$ and $0.05\pm0.15$~Myr, respectively, with isochronous estimates of their age, 1--2~Myr, found in the work of \cite{Galli2021}. We concluded that it is quite possible that the clusters Cha\,I and Cha\,II may originate from the same parent cloud.

4). The kinematics of the cluster Cha\,I-north by 40 stars and Cha\,I-south by 36 stars are analyzed in detail. Based on the trigonometric parallaxes of Gaia\,DR3, the average distance to Cha\,I-north is determined, which is ${\overline r}=191.8\pm0.4$~pc, and to Cha\,I-north~--- ${\overline r}=189.1\pm0.7$~pc. The time points at which the size of these clusters was minimal in the past have also been found. For Cha\,I-north, this moment was $t=-0.55\pm0.24$~Myr, and for Cha\,I-south $t=-0.04\pm0.18$~Myr. It is shown that approximately 1.5~Myr ago, the distance between the trajectories of these clusters was minimal.

\medskip
The authors are grateful to the reviewer for useful comments that contributed to the improvement of the work.


\begin{thebibliography}{99}

 \bibitem[\protect\citeauthoryear{Blaauw}{1946}]{Blaauw1946}
Blaauw, A., 1946, Publ. Kapteyn Astron. Lab. Groningen 52, 1

 \bibitem[\protect\citeauthoryear{Blaauw}{1964}]{Blaauw1964}
Blaauw, A., 1964, ARA\&A 2, 213

 \bibitem[\protect\citeauthoryear{Bobylev \& Bajkova}{2007}]{BobBajk2007}
Bobylev, V.V., and Bajkova, A.T., 2007, Astron. Lett. 33, 571

 \bibitem[\protect\citeauthoryear{Bobylev \& Bajkova}{2016}]{BobBajk2016}
Bobylev, V.V., and Bajkova, A.T., 2016, Astron. Lett. 42, 1 

 \bibitem[\protect\citeauthoryear{Bobylev \& Bajkova}{2020}]{BobBajk2020}
Bobylev, V.V., and Bajkova, A.T., 2020, Astron. Rep. 64, 326

 \bibitem[\protect\citeauthoryear{Bobylev \& Bajkova}{2023}]{BobBajk2023}
Bobylev, V.V., and Bajkova, A.T., 2023, Astron. Lett. 49, 410

 \bibitem[\protect\citeauthoryear{Bobylev \& Bajkova}{2024}]{BobBajk2024}
Bobylev, V.V., and Bajkova, A.T., 2024, Astron. Rep., in press

 \bibitem[\protect\citeauthoryear{Bruijne}{1999}]{Bruijne1999}
de Bruijne, J.H.J., 1999, MNRAS 310, 585

 \bibitem[\protect\citeauthoryear{Donlon II et al.}{2024}]{Donlon2024}
Donlon II, T., Chakrabarti, S., Widrow, L.M., et al., 2024, eprint  arXiv: 2401.15808

 \bibitem[\protect\citeauthoryear{Fern\'andez et al.}{2008}]{Fernandez2008}
Fern\'andez, D., Figueras, F., and Torra, J., A\&A 480, 735 (2008).

 \bibitem[\protect\citeauthoryear{Fuchs et al.}{2006}]{Fuchs2006}
Fuchs, B.,  Breitschwerdt, D., Avilez, M.A., et al., 2006, MNRAS 373, 993

 \bibitem[\protect\citeauthoryear{Galli et al.}{2021}]{Galli2021}
Galli, P.A.B., Bouy, H., Olivares, J., et al., 2021, A\&A 646, A46

 \bibitem[\protect\citeauthoryear{Gaia Collab.}{2016}]{Gaia Collab2016}
Gaia Collaboration (Prusti, T., et al.), 2016, A\&A 595, A1

 \bibitem[\protect\citeauthoryear{Gaia Collab.}{2018}]{Gaia Collab2018}
Gaia Collaboration (Brown, A.G.A., et al.), 2018, A\&A 616, 1 

 \bibitem[\protect\citeauthoryear{Gaia Collab.}{2022}]{Gaia Collab2022}
Gaia Collaboration (Vallenari, A., et al.), 2022, arXiv: 2208.0021

 \bibitem[\protect\citeauthoryear{de Geus et al.}{1989}]{Geus1989}
de Geus, E.J., de Zeeuw, P.T., and Lub, J., 1989, A\&A 216, 44

 \bibitem[\protect\citeauthoryear{de Geus}{1992}]{Geus1992}
de Geus, E.J., 1992, A\&A 262, 258

 \bibitem[\protect\citeauthoryear{Holmberg \& Flinn}{2004}]{Holmberg2004}
Holmberg, J., and Flinn, C., 2004, MNRAS 352, 440

 \bibitem[\protect\citeauthoryear{Hipparcos}{1997}]{HIP1997}
The HIPPARCOS and Tycho Catalogues, 1997, ESA SP--1200

 \bibitem[\protect\citeauthoryear{Kharchenko et al.}{2009}]{Kharchenko2009}
Kharchenko, N.V., Berczik, P., Petrov, M.I., et al., 2009, A\&A 495, 807

 \bibitem[\protect\citeauthoryear{Krisanova et al.}{2020}]{Krisanova2020}
Krisanova, O.I., Bobylev, V.V., and Bajkova, A.T., 2020, Astron. Lett. 46, 370

 \bibitem[\protect\citeauthoryear{Lindblad}{1927}]{Lindblad1927}
Lindblad, B., 1927, Arkiv f\"or Mat. Astron. och Fysik, Bd. 20, A, No~17

 \bibitem[\protect\citeauthoryear{Meingast et al.}{2021}]{Meingast2021}
Meingast, S, Alves, J., and Rottensteiner, A., 2021, A\&A 645, A84

 \bibitem[\protect\citeauthoryear{Piskunov et al.}{2008}]{Piskunov2008}
Piskunov, A.E., Schilbach, E., Kharchenko, N.V., et al., 2008, A\&A 477, 165

 \bibitem[\protect\citeauthoryear{Pecaut et al.}{2012}]{Pecaut2012}
Pecaut, M.J., Mamajek, E.E., and Bubar, E.J., 2012, AJ 74, 154

 \bibitem[\protect\citeauthoryear{Pecaut \& Mamajek}{2016}]{Pecaut2016}
Pecaut, M.J., and Mamajek, E.E., 2016, MNRAS 461, 794

 \bibitem[\protect\citeauthoryear{P\"oppel et al.}{2010}]{Poppel2010}
P\"oppel, W.G.L., Bajaja, E., Arnal, E.M., and Morras, R., 2010, A\&A 512, A83

 \bibitem[\protect\citeauthoryear{Preibish \& Zinnecker}{1999}]{Preibish1999}
Preibish, T., and Zinnecker, H., 1999, AJ 117, 2381

 \bibitem[\protect\citeauthoryear{Ratzenb\"ock et al.}{2023}]{Ratzenbock2023}
Ratzenb\"ock, S., Gro{\ss}schedl, J.E., Alves, J., et al., 2023, A\&A 678, A71

 \bibitem[\protect\citeauthoryear{Rizzuto et al.}{2011}]{Rizzuto2011}
Rizzuto, A.C.,  Ireland, M.J., and Robertson, J.G., 2011, MNRAS 416, 3108

 \bibitem[\protect\citeauthoryear{Williams et al.}{2000}]{Williams2000}
Williams, J.P., Blitz, L., and McKee, C.F.,
{\it Protostars and Planets IV} (Book -- Tucson: University of Arizona Press; eds Mannings, V., Boss, A.P., Russell, S.S., p. 9, 2000)

 \bibitem[\protect\citeauthoryear{Sartori et al.}{2003}]{Sartori2003}
Sartori, M.J., L\'epine, J.R.D., and Dias, W.S., 2003, A\&A 404, 913

 \bibitem[\protect\citeauthoryear{Sch\"onrich et al.}{2010}]{Schonrich2010}
Sch\"onrich, R., Binney, J., and Dehnen, W., 2010, MNRAS 403, 1829

 \bibitem[\protect\citeauthoryear{de Zeeuw et al.}{1999}]{Zeeuw1999}
de Zeeuw, P.T., Hoogerwerf, R., de Bruijne, et al., 1999, AJ 117, 354

 \bibitem[\protect\citeauthoryear{Zhou et al.}{2024}]{Zhou2024}
Zhou, J.-X., Li, G.-X., and Chen, B.-Q., 2024, eprint arXiv: 2402.02393

 \end{thebibliography}
 \end{document}